# Interpreting the candidate Galactic microlensing events

## Eamonn J. Kerins

*Astronomy Unit, School of Mathematical Sciences, Queen Mary & Westfield College, Mile End Road, London E1 4NS, UK.*

21 March 1995

**ABSTRACT**

Four microlensing collaborations are presently searching for compact matter in the Galaxy and all have detected possible candidates. Using the detection efficiencies recently published by the MACHO and OGLE collaborations, we present Monte-Carlo calculations of the expected optical depth, rates and timescales, along the LMC and Galactic bulge lines of sight, for dark matter in a four-component 'standard Galaxy' model with a spherically-symmetric halo and spheroid. Using the typically observed event durations we show that, whilst the halo fraction comprised of compact matter is likely to be $f_h < 0.4$, a 'no halo compact matter' hypothesis is ruled out at greater than the 80% confidence level, unless the LMC itself has a substantial halo of such objects. On the basis of the timescales observed by OGLE towards the bulge we find the rate predicted by the model to be in good agreement with the number of OGLE detections. We compute lens mass probability distributions for the various components and compare these estimates with current observational and theoretical constraints on the mass scale of baryonic dark matter. We assess the uniformity of the amplification distributions for the published EROS, MACHO and OGLE events and find that they are quite consistent with the microlensing hypothesis, although the OGLE candidate selection criteria mean that its data are particularly sensitive to photometric selection effects. The EROS team has recently placed strong limits on the density contribution of very low mass halo objects from their short timescale CCD search. On the basis of a recent study of the flux amplification of finite-size sources by Simmons, Newsam & Willis (1995) we suggest that EROS may have detected up to 5 low-amplification events due to halo lenses with mass $m \sim 10^{-7}$ M$_\odot$.

## 1 INTRODUCTION

The dark matter problem in our own and other galaxies is now well established; e.g. Rubin & Ford (1970), Ostriker & Peebles (1974), Ashman (1992). The inferred mass of galactic dark matter is typically estimated to be an order of magnitude more than that of the luminous parts of galaxies. Many possible solutions have been invoked from the minimal assumption of dark baryonic matter (Carr 1994) to the existence of a new form of weakly interacting matter [often referred to as cold dark matter (CDM) or weakly interacting massive particles (WIMPs)].

It is known from the observed abundances of the light elements that baryonic matter cannot provide more than 10% of the closure density of the Universe (Walker et al. 1991), but this still admits the possibility that it could provide all or some of the galactic dark matter. Indeed, the same observations imply a lower limit for the baryon density which is three times greater than the density in visible baryons (Persic & Salucci 1992) and so require some of the dark matter to be in baryonic form. Tentative measurements of a high deuterium abundance in a high-redshift primordial gas cloud by Songaila et al. (1994) and Carswell et al. (1994) imply a baryon density which is comparable to the visible density, but the discrepancy with solar system abundances implies that these observations may be due to an intervening hydrogen cloud at lower redshift (Steigman 1994). Subsequently, observations of another high-redshift cloud have inferred a much lower deuterium abundance (Carswell 1994). A comparison of the amount of gas present in some galaxy clusters with predictions from numerical simulations implies that the contribution of baryons to the total cluster mass may exceed the amount compatible with standard nucleosynthesis calculations (White et al. 1993), unless the density of the universe is only around 20% of the closure density. It therefore seems likely that baryonic matter is relevant on galactic scales.

The nature of such baryonic matter is unknown, but there are a number of observational and theoretical constraints on many of the possible candidates; e.g. Carr (1994), Dalcanton et al. (1994), Silk (1994). Paczyński (1986) suggested that microlensing could be used to look for compact dark matter in the Galactic halo by simultaneously mon-itoring the brightness of $10^{6-7}$ stars for a few years for signs of variations indicative of the microlensing effect. Four experiments are now in progress: MACHO (Alcock et al. 1993); EROS (Aubourg et al. 1993); OGLE (Udalski et al. 1993); and DUO (D. Valls-Gabaud, private communication). To date these teams have detected more than 70 candidate events.

The plan of this paper is as follows. In Section 2 we discuss the lensing properties of dark matter in a 'standard Galaxy' model comprising a spherically-symmetric halo, thin and thick discs and a spherically-symmetric spheroid/bulge component. We use the recent efficiency estimates from the MACHO and OGLE collaborations and the typically observed event durations to compare the predictions of this model to the observations, and assess to what extent the observations disagree with the model predictions. We also discuss the implications of the current results from EROS short-timescale CCD search which is sensitive to very low mass objects. In Section 3 we calculate mass probability distributions for the lenses along both the LMC and bulge lines of sight and discuss their consistency with other theoretical and observational constraints on the mass scale baryonic dark matter. We also test the distribution of event amplifications observed by EROS, MACHO and OGLE against the expected distribution for microlensing events. Section 4 discusses the results and implications of the previous sections.

## 2    OPTICAL DEPTH AND LENSING RATE TOWARDS THE LMC AND BULGE

The number of lensing events in progress at a particular instant in time is given by the optical depth:

$$\tau = \frac{\pi}{N_s} \int \frac{R_e^2(m,x,L)}{m} \rho_l(x,l,b) \frac{d^3 N_s}{dl\, db\, dL}\, dx\, dl\, db\, dL, \quad (2.1)$$

where the integral sign implies integration over the source distance $L$ (measured from the observer), lens distance $x < L$, Galactic longitude $l$ and latitude $b$. $\rho_l$ denotes the lens density, $m$ the lens mass, $N_s$ the number of sources monitored and $R_e = [4Gmx(L-x)/c^2L]^{1/2}$ the Einstein radius, which is the radius of the image ring formed (as measured along the lens plane) when the lens is perfectly aligned with the observer–source line of sight. If the lens and source are separated by a distance $uR_e$ then the source brightness is amplified by a factor

$$A = \frac{u^2 + 2}{u(u^2 + 4)^{1/2}}. \quad (2.2)$$

$R_e$ provides a natural lensing length scale and from here on we shall define an event to be where the lens–source separation is less than $R_e$ [i.e. $u < 1$ which, from eqn (2.2), implies $A > 1.34$].

The number of events due to lenses of mass $m$ per unit time interval is the lensing rate:

$$\Gamma = \frac{2}{N_s} \int \frac{R_e(m,x,L)}{m} \rho_l(x,l,b) \langle V_t(x,L) \rangle_\theta$$
$$\times \frac{d^3 N_s}{dl\, db\, dL}\, dx\, dl\, db\, dL, \quad (2.3)$$

where $V_t$ is the relative tangential velocity of the lens with respect to the line of sight. $V_t$ depends on the respective motions of the lens and source and their relative orientation $\theta$ (Griest 1991; Kiraga & Paczyński 1994) and so there is an implied summation over all possible source and lens velocities, as well as over $l$, $b$, $L$ and $x < L$. The average event timescale is $\langle t_e \rangle = \langle R_e/V_t \rangle = 2\tau/\pi\Gamma$ and is thus determined by the optical depth and lensing rate. Note that we are here defining the event timescale as the Einstein-radius crossing time. The MACHO collaboration adopt the Einstein-diameter crossing time as their definition and so all timescales in this paper need to be multiplied by 2 before being compared to the MACHO results.

There are many possible locations for the lensing events. Towards the LMC possible sources include the Galactic halo (Paczyński 1986; Griest 1991; De Rújula, Jetzer & Massó 1991); the thin or thick discs (Paczyński 1991; Griest et al. 1991; Gould 1994a; De Rújula, Giudice, Mollerach & Roulet 1994); the spheroid (Giudice, Mollerach & Roulet 1994); and the LMC itself (Gould 1993; Sahu 1994; De Rújula, Giudice, Mollerach & Roulet 1994). The halo, disc and spheroid may also contribute to lensing towards the Galactic bulge, along with the bulge itself (Kiraga & Paczyński 1994; Evans 1994; De Rújula, Giudice, Mollerach & Roulet 1994).

Gould, Miralda-Escudé & Bahcall (1994) have investigated the relative contributions of the halo, thin and thick disc components to the optical depth towards the LMC, SMC and bulge. A similar analysis of the lensing rates could in principle provide a more critical comparison, for a particular Galactic model, since, not only does the rate depend on the lens density distribution, it also depends on the lens mass distribution and the lens and source velocity profiles. However, a naive analysis which assumes that events of all timescales are uniformly sampled by the microlensing experiments is likely to be misleading since, in practice, sampling and photometric selection effects mean that microlensing observations do not uniformly sample the lensing rate. The frequency of individual observations dictates the minimum observable timescale $t_e(\min)$, whilst the total duration of the experiment determines the maximum observable event duration $t_e(\max)$. Timescales between $t_e(\min)$ and $t_e(\max)$ are sampled with an efficiency $\epsilon(t_e) \leq 1$ (Griest 1991). Photometric effects, such as poor seeing or excessive 'crowding' of the stellar images, also mean that some event durations will be observed in preference to others.

The different density and velocity profiles of the various components give rise to different characteristic lensing event durations, and so the effect of the sampling and photometric efficiencies can be to enhance the observed rate of lenses from one component relative to another. The MACHO and OGLE collaborations have recently published their detection efficiencies (Bennett et al. 1995; Alcock et al. 1995; Udalski et al. 1994) and so we shall use these in examining the relative contributions of a four-component Galactic model to the observed rate and optical depth along the LMC and bulge lines of sight.

In evaluating the lensing rate towards the LMC we will neglect the contribution due to lenses in the LMC itself and consider only Galactic sources, although we shall discuss the likely LMC contribution. In the case of lensing towards the bulge we follow Kiraga & Paczyński (1994) in taking account of the line-of-sight depth of the bulge sources.

## 2.1 Galactic model

The three main components comprising our Galaxy are the rotationally-supported disc, the partially rotationally-supported visible halo and the dark outer halo. There is some evidence that the disc comprises both a 'thin' and 'thick' component, whilst the visible halo can be sub-divided into a central bulge surrounded by a spheroidal component. [For reviews see Bahcall (1986) and Gilmore (1989).] We shall utilise simple analytic models for these components. For the dark halo we shall adopt the softened-isothermal sphere of Bahcall & Soneira (1980):

$$\rho(\text{halo}) = \rho_0 \left( \frac{a_h^2 + R_0^2}{x^2 - 2xR_0 \cos b \cos l + a_h^2 + R_0^2} \right) \quad (2.4)$$

where $\rho_0 = 0.01$ $M_\odot$ pc$^{-3}$ is the local density, $R_0 = 8$ kpc is the Sun's Galactocentric distance and $a_h \simeq 5$ kpc is the halo core radius. We assume that the halo lenses have an isotropic random velocity distribution (as measured in the rest frame of the Galaxy) which, in two dimensions, resembles the Rayleigh distribution:

$$P(V_l)\, dV_l = 2\,(V_l/V_c)\exp[-(V_l/V_c)^2]\,(dV_l/V_c), \quad (2.5)$$

where $V_c = 220$ km s$^{-1}$. The optical depth and lensing rate produced by a halo of the form of eqns (2.4) and (2.5) has been calculated by Griest (1991) and has recently been constrained by the MACHO collaboration (Alcock et al. 1995). Here, we shall take account of other possible sources of lensing in assessing the contribution of the halo.

For the thin and thick discs we take the density distribution to be

$$\rho(\text{disc}) = \rho_0$$
$$\times \exp\left\{ \left[ \frac{R_0 - (x^2 \cos^2 b - 2xR_0 \cos b \cos l + R_0^2)^{1/2}}{h} \right] \right.$$
$$\left. - \left[ \frac{x \sin |b|}{H} \right] \right\}, \quad (2.6)$$

where $h$ and $H$ are the disc scale length and scale height, respectively, and $\rho_0 = \rho_{vis} + \rho_{dark}$, with $\rho_{vis}$ the contribution of the visible matter and $\rho_{dark}$ the contribution of the dark matter. For the thin disc we take $\rho_0 = 0.2$ $M_\odot$ pc$^{-3}$, with $\rho_{dark} = 0.1$ $M_\odot$ pc$^{-3}$ and $H = 0.3$ kpc, whilst for the thick disc we assume $\rho_0 = 4 \times 10^{-3}$ $M_\odot$ pc$^{-3}$, with $\rho_{dark} = 2 \times 10^{-3}$ $M_\odot$ pc$^{-3}$ and $H = 1.0$ kpc. We adopt $h = 3.5$ kpc for both components. We assume that both components are rotationally supported with $V_{rot} = 220$ km s$^{-1}$ for the thin disc and $V_{rot} = 170$ km s$^{-1}$ for the thick component. In addition, we take both discs to have a random velocity component of the form of eqn (2.5) with $V_c = 30$ km s$^{-1}$ and 60 km s$^{-1}$ for the thin and thick disc, respectively.

For the spheroid and bulge (which, form here onwards, will be collectively referred to as the bulge) we adopt the simple spherically-symmetric model used by De Rújula et al. (1994) where

$$\rho(\text{bulge}) = \rho_0 \left[ \frac{a_b^{1/2} + R_0^{1/2}}{a_b^{1/2} + (x^2 - 2xR_0 \cos b \cos l + R_0^2)^{1/4}} \right]^{-7} \quad (2.7)$$

with $\rho_0 = \rho_{vis} + \rho_{dark}$. There is some uncertainty as to the amount of dark matter in the bulge and so we shall discuss two scenarios: $\rho_0 = 10^{-3}$ $M_\odot$ pc$^{-3}$, with $\rho_{dark} = 8.75 \times 10^{-4}$ $M_\odot$ pc$^{-3}$ (which, from here on, will be referred to as the 'dark matter-dominated' bulge), and $\rho_0 = 2.5 \times 10^{-4}$ $M_\odot$ pc$^{-3}$, with $\rho_{dark} = 1.25 \times 10^{-4}$ $M_\odot$ pc$^{-3}$. The core radius is taken to be $a_b = 0.17$ kpc. We take the bulge to have a rotational motion $V_{rot} = 100$ km s$^{-1}$ and a random component given by eqn (2.5) with $V_c = 100$ km s$^{-1}$. Spherical symmetry is likely to be an over simplistic assumption since much of the observational evidence points towards a flattened or perhaps bar-like bulge [e.g. Blitz & Spergel (1991); Weiland et al. (1994)]. However, calculations based on spherical-symmetry at least provide a 'bench mark' against which any asymmetry can be quantified. One of the aims of this paper is to evaluate to what extent microlensing observations provide evidence for bulge asymmetry.

The visible component of the local spheroid density is normalised to 1/800th that of the thin disc, and the visible component of the thick disc density to 1/50th. The combined column density (dark and visible matter) of the four components within 1.1 kpc of the Galactic plane is 83 $M_\odot$ pc$^{-2}$, consistent with observational constraints (Bahcall, Flynn & Gould 1992). The total column density in the thin and thick discs is 64 $M_\odot$ pc$^{-2}$, which is also consistent with rotation curve constraints. We shall refer to this as the 'standard model', since it is very much along the lines of the Galactic models advocated by Bahcall & Soneira (1980) and by Kuijken & Gilmore (1989). The main uncertainty regarding the standard model concerns the existence of the thick disc. However, it turns out that the lensing contribution of the thick disc is anyway rather small.

We are concerned here with the lensing properties of the dark matter components of this model. The visible components also make some contribution to the lensing rate, although the number of achromatic lensing events (which generally require the lens to be invisible) is small. This fact is reinforced by recent limits on the lensing contribution of M-dwarfs (Bahcall, Flynn, Gould & Kirhakos 1994; Hu, Huang, Gilmore & Cowie 1994).

## 2.2 Source distribution

As well as specifying the lens distribution, we also need to do the same for the sources. In the case of lensing towards the LMC, the sources can be assumed to be all at the same distance, since the distance to the LMC is significantly greater than its line-of-sight depth. The angular distribution of the sources is also unimportant. We shall therefore adopt $L = 50$ kpc, $l = 280°$ and $b = -33°$ for the LMC sources. We shall, however, follow Griest (1991) in taking account of the tangential motion of the LMC.

In calculating the optical depth and lensing rate towards the bulge, the line-of-sight depth can provide an important correction (Kiraga & Paczyński 1994). We assume that all sources reside in the bulge although one should also expect a small contribution to the lensing rate from sources in the disc. Taking the number density of sources per unit V-band luminosity to be $dn_s/d\mathcal{L} \propto \mathcal{L}^{-\beta}$ between $\mathcal{L}_l$ and $\mathcal{L}_u$, the variation of the number of sources with $l$, $b$ and $L$ is given by

$$d^3 N_s \propto [\mathcal{L}_u^{1-\beta} - \mathcal{L}_{min}^{1-\beta}(L)]\rho_s(l,b,L)L^2 \cos b\, dl\, db\, dL$$
$$(\beta \neq 1), \quad (2.8)$$

where the source density $\rho_s$ is given by eqn (2.7) and $\mathcal{L}_{min} = \max[\mathcal{L}_l, K(L/\text{kpc})^2]$ is the minimum observable luminosity, with $K = 10^{5.9-0.4V(lim)}\mathcal{L}_\odot$ (neglecting extinction) and $V(lim) \simeq +21$ mag the limiting apparent V-band magnitude of the observation. For our calculations we adopt $\beta = 2$. Since $\mathcal{L}_{min} \ll \mathcal{L}_u$ the maximum values of $L$ and $\mathcal{L}_u$, eqn (2.8) turns out to be insensitive to $\mathcal{L}_u$ for $\beta > 1$. Furthermore, assuming $\mathcal{L}_l < 10^{-3}\,\mathcal{L}_\odot$ (corresponding to a source mass $m_s < 0.1\,M_\odot$), $\mathcal{L}_{min} > \mathcal{L}_l$ everywhere along the line of sight except very close to the observer ($L < 0.5$ kpc). This means that eqn (2.8) is also insensitive to the precise value of $\mathcal{L}_l$ and so $d^3N_s \propto \rho_s(L)L^{2(2-\beta)} = \rho_s(L)$.

## 2.3   Results

To evaluate eqns (2.1) and (2.3) we used Monte-Carlo integration. The (Gaussian) error estimate was typically less than 1% for $10^6$ evaluations. In reality, the error distribution is not Gaussian, although we do not expect the actual error to be more than a few percent. The results are given for the LMC and bulge (Baade's Window) in Tables 1 and 2, respectively.

The first rows of Tables 1 and 2 give the optical depth, lensing rate and average timescale, under the assumption that the detection efficiency $\epsilon(t_e) = 1$ for all timescales, whilst for the other rows detection efficiencies have been included in the calculation. For the LMC line of sight we use the MACHO efficiencies (Alcock et al. 1995), and for the bulge line-of-sight we use the OGLE efficiency estimates (Udalski et al. 1994). The MACHO collaboration has also published efficiency estimates for its bulge observations (Bennett et al. 1995), although they are only sampling efficiencies and do not take account of photometric effects. We shall therefore only compare the predictions of our Galactic model with the OGLE observations.

### 2.3.1   Lensing towards the LMC

The first thing to notice from Table 1 is that the combined optical depth of the thin disc, thick disc and bulge accounts for only 4% of the total optical depth for our adopted model. The halo is by far the most dominant source of Galactic lensing. The lensing rate is likewise dominated by the halo, with the combined bulge and disc contribution ranging from 1% for 1 $M_\odot$ deflectors up to 9% for $10^{-4}\,M_\odot$ lenses (assuming all components are comprised of lenses of a similar mass). If detection efficiencies are neglected the contribution of the disc and bulge to the total rate is 2.4% and so including the efficiency estimates results in an enhancement of the relative contribution of the bulge and disc to the overall rate for lenses below 0.1 $M_\odot$, and a reduction above 0.1 $M_\odot$.

The average observed timescale is essentially given by the average timescale for the halo component. The average timescale for the 3 events observed in the first year MACHO data is 13.8 d (Alcock et al. 1995) which, from Table 1 (and Section 3.1), suggests halo lenses of around 0.01 $M_\odot - 0.1\,M_\odot$. Taking the observed timescales as a guideline, it is an interesting exercise to estimate the maximum contribution to the total rate from the bulge and disc components, allowing their lens masses to be different from those in the halo, and assuming all components are completely comprised of compact objects. The halo rate peaks for $m \sim 0.01\,M_\odot$, although the observed timescales are quite compatible with 0.1 $M_\odot$ deflectors, which give only half the rate (around 16 events per $10^7$ stars per year). The bulge and disc rates all peak for $m \sim 10^{-3}\,M_\odot$ lenses with a combined total of around 1.3 events per $10^7$ stars per year. This implies a contribution of 8% to the total rate. The contribution is as much as 13% for a dark matter-dominated bulge.

The MACHO exposure of $9.7 \times 10^6$ star-years (8.6 million stars monitored over 1.1 years) actually yielded only 3 events (Alcock et al. 1995), rather than the 16 or so events predicted by the above model. The relative likelihood of only 3 events being detected, when the expectation is 16, is only $3 \times 10^{-4}$. Turning this around: given 3 events have been observed, the 68% confidence level (CL) lower limit on the expectation value for the rate is 2.9 events per year (1.4 at the 95% CL) and the 68% CL upper limit is 4.6 (7.7 at the 95% CL). Fixing the combined disc and bulge rate at 1.3 events per year, the contribution of the halo is anything between 55% (7%) and 72% (83%) of the total rate at the 68% CL (95% CL). Translating this into an estimate of the fraction of our adopted halo comprised of compact matter, we find the halo fraction $f_h$ to lie between $0.10 < f_h < 0.21$ (68% CL) or $0.006 < f_h < 0.40$ (95% CL), assuming halo lenses to have a mass of around 0.1 $M_\odot$. For a dark matter-dominated bulge the contribution of the disc and bulge is as much as 2.3 events per year (assuming $10^{-3}\,M_\odot$ deflectors) and so the inferred halo fraction is $0.04 < f_h < 0.14$ (68% CL) or $0 \le f_h < 0.34$ (95% CL). It therefore seems clear that the halo is not completely comprised of compact matter, although the 'no halo compact matter' hypothesis is ruled out at the 96% CL (or at the 80% CL in the case of a dark matter-dominated bulge).

In fact, there a number of reasons why the combined disc and bulge rate would not be as high as quoted above. Firstly, there is evidence that sub-stellar objects (or brown dwarfs) probably do not dominate the density in either the thin disc (Kroupa, Tout & Gilmore 1993; Kerins & Carr 1994) or the bulge (Bahcall, Flynn, Gould & Kirhakos 1994), implying that the bulk of these lenses must be larger than 0.1 $M_\odot$ (i.e. stellar remnants or hydrogen-burning stars) and thus have a lower lensing rate. In the case of hydrogen-burning stars, the lenses cannot be nearby, otherwise they would be directly observable and would not, in general, give rise to achromatic lensing events (since the lens and source would not, in general, have the same colour). Along the LMC line of sight, the disc and bulge density distributions peak close to the observer, so there is a significant contribution to the rate from visible lenses which in general produce non-achromatic light curves. For this reason we do not expect the visible components of the disc and bulge to give rise to many achromatic lensing events towards the LMC. These reductions may, in part, be offset by the rate due to lenses in the LMC itself, which is not considered here [see e.g. De Rújula et al. (1994)]. The LMC may make a substantial contribution to the rate if it has its own dark halo of compact matter, although a scenario in which the LMC halo is completely comprised of compact matter, but our own halo is not, seems contrived. In the absence of such a halo, the contribution of the LMC (from lenses in the LMC disc) would be at most comparable to our own disc.

**Table 1.** The optical depth, lensing rate and average timescale for events towards the LMC ($l = 280°, b = -33°$). The first row shows these quantities under the assumption of perfectly efficient detection ($\epsilon = 1$) and a lens mass of 1 $M_\odot$. For other masses $m$ these values scale as $(m/M_\odot)^{-1/2}$ for $\Gamma$ and $(m/M_\odot)^{1/2}$ for $\langle t_e \rangle$. The other rows incorporate the efficiency estimates of the MACHO collaboration (Alcock et al. 1995). $\tau$ and $\Gamma$ are displayed as the sum of the individual component contributions (bulge + thin disc + thick disc + halo) with their combined total in brackets. $\langle t_e \rangle$ is similarly displayed with the rate-weighted average appearing in brackets. For the dark matter-dominated bulge, the bulge optical depth and lensing rate values tabulated below should be multiplied by a factor of 7.

| $\log(m/M_\odot)$ | $\tau/10^{-8}$ | $\Gamma/(10^{-7}$ events star$^{-1}$ yr$^{-1}$) | | $\langle t_e \rangle$/days | |
|---|---|---|---|---|---|
| 0 ($\epsilon = 1$) | $0.33 + 1.84 + 0.35 + 56$ (59) | $0.08 + 0.35 + 0.09 + 21$ | (21) | 96, 120, 95, 63 | (64) |
| 0 | as above | $0.01 + 0.03 + 0.01 + 4.2$ | (4.3) | 64, 75, 65, 46 | (46) |
| -1 | as above | $0.07 + 0.29 + 0.07 + 16$ | (16) | 28, 34, 28, 20 | (20) |
| -2 | as above | $0.14 + 0.72 + 0.16 + 28$ | (29) | 11, 13, 11, 8.0 | (8.2) |
| -3 | as above | $0.17 + 0.97 + 0.18 + 24$ | (25) | 4.4, 5.1, 4.2, 3.5 | (3.5) |
| -4 | as above | $0.06 + 0.47 + 0.06 + 5.8$ | (6.4) | 2.5, 2.6, 2.4, 2.5 | (2.5) |

**Table 2.** The optical depth, lensing rate and average timescale for events towards the bulge ($l = 0.°7$ to $1.°3$, $b = -4.°2$ to $-3.°5$). The layout is analogous to Table 1 with the first row values assuming perfectly efficient detection and a lens mass of 1 $M_\odot$. The other rows incorporate the efficiency estimates of the OGLE collaboration (Udalski et al. 1994). The values assume that all sources reside in the bulge component and have a V-band luminosity function $dn/d\mathcal{L} \propto \mathcal{L}^{-2}$. For the dark matter-dominated bulge, the bulge optical depth and lensing rate values tabulated below should be multiplied by a factor of 7.

| $\log(m/M_\odot)$ | $\tau/10^{-8}$ | $\Gamma/(10^{-7}$ events star$^{-1}$ yr$^{-1}$) | | $\langle t_e \rangle$/days | |
|---|---|---|---|---|---|
| 0 ($\epsilon = 1$) | $5.5 + 84 + 3.5 + 15$ (108) | $3.4 + 38 + 1.5 + 13$ | (56) | 37, 51, 55, 27 | (45) |
| 1 | as above | $0.35 + 2.7 + 0.10 + 1.6$ | (4.8) | 75, 92, 90, 67 | (82) |
| 0 | as above | $1.4 + 16 + 0.61 + 5.0$ | (23) | 36, 47, 49, 28 | (42) |
| -1 | as above | $3.1 + 41 + 1.6 + 9.9$ | (56) | 14, 18, 19, 10 | (16) |
| -2 | as above | $4.4 + 66 + 2.6 + 12$ | (85) | 5.5, 6.5, 7.3, 4.0 | (6.1) |
| -3 | as above | $2.3 + 48 + 2.1 + 3.5$ | (56) | 3.1, 3.1, 3.5, 2.8 | (3.1) |
| -4 | as above | $0.36 + 7.0 + 0.4 + 0.39$ | (8.2) | 2.7, 2.6, 2.8, 2.7 | (2.6) |

### 2.3.2 EROS short timescale search

The EROS collaboration is undertaking two microlensing programs. One is a long-timescale search towards the LMC and SMC, which is sensitive to event timescales $t_e \gtrsim 2$ days and has so far uncovered 2 LMC events (Aubourg et al. 1993). The other is a high time resolution CCD search towards the bulge, which is sensitive to events of 30 min $\lesssim t_e \lesssim 7$ days duration, roughly corresponding to lenses of mass $10^{-8}$ $M_\odot \lesssim m \lesssim 10^{-2}$ $M_\odot$ (Aubourg et al. 1995). The collaboration has detected no events with $A > 1.2$ having monitored $8.2 \times 10^4$ stars for 10 months, thus ruling out objects with $5 \times 10^{-8}$ $M_\odot < m < 7 \times 10^{-4}$ $M_\odot$ from providing all of the halo dark matter (90% CL). Interestingly, they did find 5 candidate events with $1 < A < 1.2$, although these are below their adopted amplification threshold.

Simmons, Newsam & Willis (1995) have calculated the flux amplification for sources with radius comparable to or greater than the Einstein radius of the lens (as projected onto the source plane). They find that, for a given amplification $A$, there is a critical source radius $R_s = R_c \simeq 3 R_e L/x$ at which the impact parameter $u$ decreases rapidly as $R_s$ increases, until the source can no longer be amplified by as much as $A$ for any value of $u$ [see Fig. 6 of Simmons, Newsam & Willis (1995)]. For halo lenses

$R_c \sim 30$ $R_\odot$ $(m_l/10^{-6}$ $M_\odot)^{1/2}$, assuming $x \sim 8$ kpc (which is a typical lens distance towards the LMC). In this case, sources with $R_s \sim 30$ $R_\odot$ can be amplified by $A \sim 1.2$ for a range of impact parameters $0 \leq u \lesssim 2$. If we take LMC main-sequence stars to have a power-law mass function $dn_s/dm_s \propto m_s^{-2.7}$ [where the exponent used is the Scalo value (Scalo 1986) for the local Solar neighbourhood, although determinations of LMC mass function appear to give similar results (Hill, Madore & Freedman 1994)], and a mass-radius relation $m_s \propto R_s^{0.85}$ to give $dn_s/dR_s \propto R_s^{-2.5}$, then the fraction of source stars with a radius sufficiently large that an intervening lens of mass $m_l$ can only amplify its brightness by $A < 1.2$ (irrespective of $u$) is

$$f(R_s > R_c) = \frac{\int_{R_c}^{R_{max}} dn_s/dR_s \, dR_s}{\int_{R_{min}}^{R_{max}} dn_s/dR_s \, dR_s}. \tag{2.9}$$

For the mass function and mass-radius relation we have adopted, eqn (2.9) is insensitive to the maximum radius $R_{max}$ and so $f \simeq (R_c/R_{min})^{-1.5}$, where $R_{min}$ is the radius corresponding to a source star of luminosity $\mathcal{L}_{min}$. For the EROS CCD survey $V(lim) \simeq +19$, so $\mathcal{L}_{min} \simeq 50$ $\mathcal{L}_\odot$ which implies $R_{min} \simeq 5$ $R_\odot$. Hence, $f \sim 0.3$ $(m_l/10^{-6}$ $M_\odot)^{-3/4}$. If one expects to detect $N$ events due to lenses of mass $m_l$ then the probability that they all lens source stars with $R_s > R_c$ and so produce an amplification $A < 1.2$ is $P(N) = f^N \sim 0.3^N (m_l/10^{-6}~M_\odot)^{-3N/4}$.

The EROS team rejects its five candidates with $A < 1.2$ on the basis that, if their amplifications are given by the point-source approximation [eqn (2.2)], they occupy an improbable portion of the impact-parameter distribution. From our crude analysis, we see that the probability that 5 lensing events all occur for source stars with $R_s > R_c$ (in which case the point-source approximation is invalid and only low-amplification events are expected) is only around $10^{-3}$ for $10^{-6}~M_\odot$ lenses but rapidly tends to unity for lens masses approaching $10^{-7}~M_\odot$. This estimate is quite crude in many respects but does seem to indicate that at least some of the 5 EROS candidates could be interpreted as true microlensing events if the lenses have a low mass, say $m_l \sim 10^{-7}~M_\odot$, and are lensing source stars with large radius ($R_s \sim 10~R_\odot$). The fact that the source stars all appear to be much brighter than average is also consistent with this idea.

In fact, only 30% of the stars monitored by EROS are main sequence, the rest are giants. Our estimate may therefore be on the pessimistic side. Intriguingly, the number of events due to $10^{-7}~M_\odot$ lenses expected by EROS is 5.6 if the halo is completely comprised of such objects!

### 2.3.3   Lensing towards the Galactic bulge

Table 2 shows the expected optical depth, lensing rates and average event times for the bulge line of sight. The actual region of the bulge corresponds to the Baade's Window area monitored by the OGLE collaboration, in which 7 of the 10 events reported by Udalski et al. (1994) were found. In the case of a non dark matter-dominated bulge, the thin disc provides the dominant contribution to the optical depth (78%), although the halo is also an important source (contributing 14%). The disc likewise dominates the rate, providing up to 85% of the total rate in the case of $10^{-4}~M_\odot$ lenses. The disc dominates the rate even for the case of a dark matter-dominated bulge, although the bulge contribution is comparable. The effect of the efficiencies is to enhance the rate of the halo and bulge, relative to the thin disc, for masses above $1~M_\odot$ but reduce their contribution below $1~M_\odot$. The opposite is true for the thick disc, the contribution of which increases relative to the thin disc for lower mass lenses. However, since the thick disc is not a major contributor to the overall rate, the thin disc provides a greater proportion of the observed total rate for lower mass lenses.

The average event timescale for the 7 OGLE lenses found in the Baade's Window region is 33 d, implying typical lens masses in the $0.1~M_\odot - 1~M_\odot$ range from Table 2 (see also Section 3.1). Of these, 2 have minimum impact parameters larger than $R_e$ ($u > 1$), leaving 5 events with an average timescale of 31 d. On the basis of the average observed timescale, the expected rate is $\sim (2.3 - 5.6) \times 10^{-6}$ events per star per year. The OGLE collaboration monitored $9.5 \times 10^5$ stars in 1992 (which yielded 5 of the 5 events) and $7.5 \times 10^5$ in 1993 (allowing for field overlaps and contamination due to disc sources) and so the expectation rate is $2.2 - 5.3$ events in 1992 and $1.7 - 4.2$ events in 1993. These rate intervals include the observed values and so our standard Galaxy is not actually at odds with the OGLE data as far as the observed rate is concerned.

However, there does appear to be a discrepancy between the model and the observations concerning the optical depth. The total optical depth for the model is only $(1.1 - 1.4) \times 10^{-6}$, where as the value of $\tau$ inferred by the OGLE collaboration (based on all of its events) is $(3.3 \pm 1.2) \times 10^{-6}$ (Udalski et al. 1994) [a similar value has been obtained by the MACHO collaboration, based on its sample of 13 'clump giant' lensing events (Bennett et al. 1995)]. It is not clear how serious this discrepancy is, since one might not be surprised by an optical depth estimate, inferred from only 10 events, deviating significantly from the expectation value. It is unclear how large a deviation is allowed (since it depends on the relative contributions of the various components), although it should be much larger than suggested by Poisson estimates.

It is interesting to note from Table 2 that, whilst the (non dark matter-dominated) bulge contributes less than 6% to the total rate (for a lens mass $m < 1~M_\odot$), the rate-weighted average timescale for all components is typically fairly close to the average timescale for the bulge. Therefore, in the case of a dark matter-dominated bulge $0.1~M_\odot - 1~M_\odot$ lenses are still favoured. However, for this mass range, bulge lenses become a major contributor to the rate with $(1.0 - 2.2) \times 10^{-6}$ events per star per year, and the predicted total rate for the OGLE 1992 and 1993 observing seasons becomes $3.0 - 7.0$ and $2.3 - 5.6$ events, respectively. Note, however, that recent HST observations of the local bulge population appear to reduce the possibility of much bulge dark matter (Bahcall, Flynn, Gould & Kirhakos 1994). We have seen that the observed rate towards the LMC implies that the halo fraction is likely to be $f_h \sim 0.2$ rather than 1. This would result in a small reduction in the overall rate for lenses between $0.1 - 1~M_\odot$ but the model would still be quite consistent with the observed rate towards the bulge. Of course, one would also expect a small contribution to the rate from the lensing of disc sources.

Alternative Galactic models invoked to reconcile the predicted and observed optical depth include a bar-like bulge (Paczyński 1994a; Evans 1994; Zhao, Spergel & Rich 1995), for which there is some independent observational support (Blitz & Spergel 1991; Weiland et al. 1994). A more radical suggestion for increasing the optical depth involves assuming a maximal disc (Gould 1994a; Alcock et al. 1994). In this scenario, almost all of the rotation curve velocity inside $R_0 = 8$ kpc is provided by the disc, thereby allowing a much larger density (and implying a substantially reduced halo). However, Paczyński et al. (1994b) claim that there is evidence of a dramatic decrease in the disc density within the inner $\sim 4$ kpc of the Galaxy. Such a hollow disc could not provide the required optical depth and so a bar-like bulge aligned close to the observer's line of sight would appear the more plausible alternative. (A hollow disc would likewise lower the rate and optical depth estimates calculated in this paper, as well as altering the average timescale estimates.) These models ought to be testable in the near future with increased angular coverage by the microlensing experiments. Whilst there is other evidence which points towards a Galactic model which differs from the standard one investigated here, we believe that, on the basis of microlensing observa-tions alone, the case for a radical change in the model is not yet an overwhelming one.

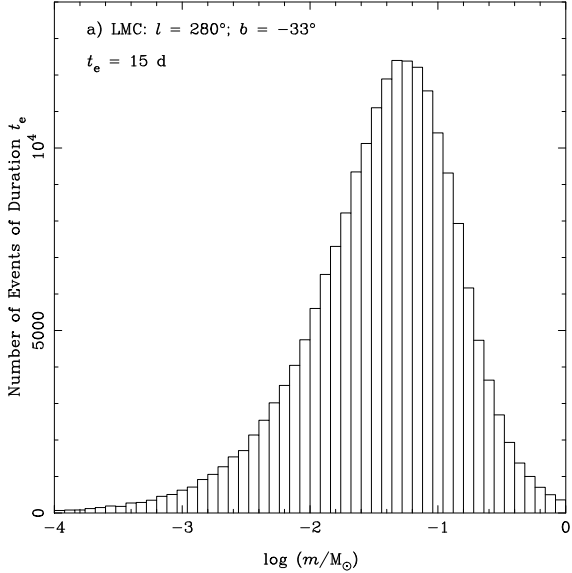

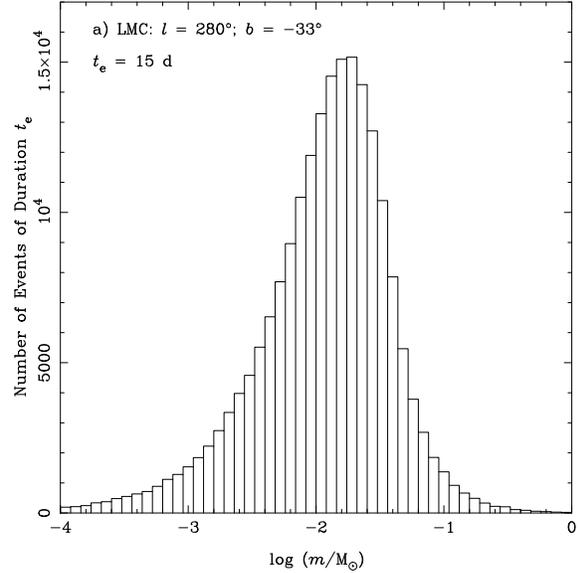

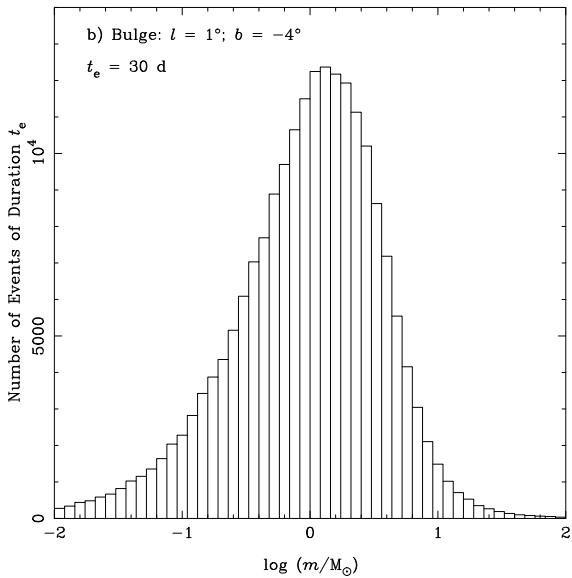

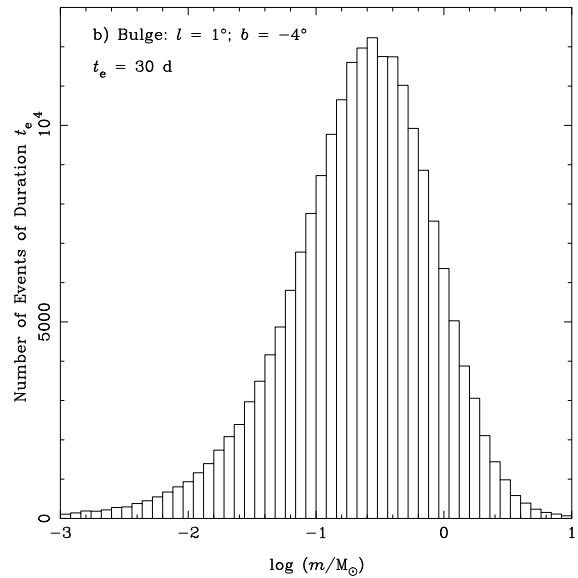

**Figure 1.** The number of simulated halo events of mass $m$ with event duration $t_e$ along the (a) LMC and (b) Galactic bulge (Baade's Window) lines of sight. The adopted values for $t_e$ reflect the event durations that are typically being observed. The most likely lens mass is $0.05\ M_\odot\ (t_e/15\ \text{d})^2$ for the LMC direction and $1.2\ M_\odot\ (t_e/30\ \text{d})^2$ for the bulge.

**Figure 2.** The number of simulated thin-disc events of mass $m$ with event duration $t_e$ along the (a) LMC and (b) Galactic bulge (Baade's Window) lines of sight. The most likely lens mass is $0.02\ M_\odot\ (t_e/15\ \text{d})^2$ for the LMC direction and $0.3\ M_\odot\ (t_e/30\ \text{d})^2$ for the bulge.

## 3  LENS MASS ESTIMATES AND EVENT AMPLIFICATIONS

Having looked at the optical depth, lensing rate and average event duration, we now turn our attention to the individual lenses themselves.

### 3.1  Lens mass

Mass estimates of individual lenses cannot be made very precisely without additional parallax (Gould 1994b), source-polarisation (Simmons, Newsam & Willis 1995) or a combination of high-precision astrometric and photometric observations (Hog, Novikov & Polnarev 1995). This is because the event duration not only depends on the lens mass but also on its relative tangential velocity, as well as the lens and

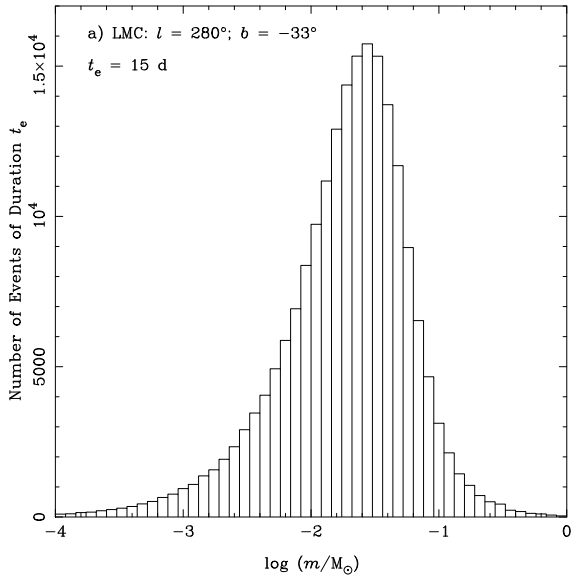

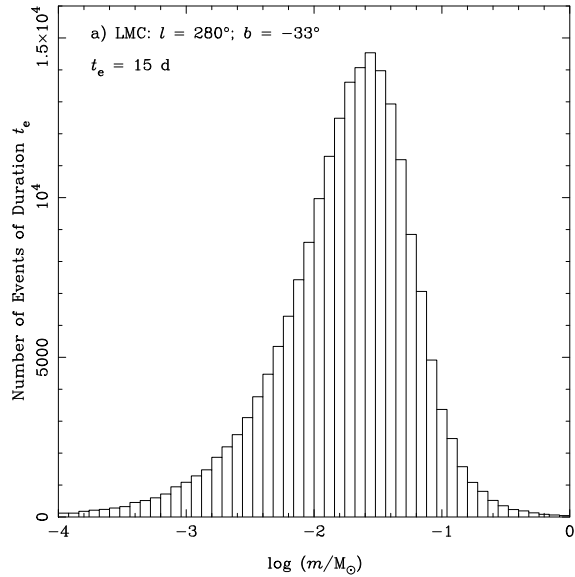

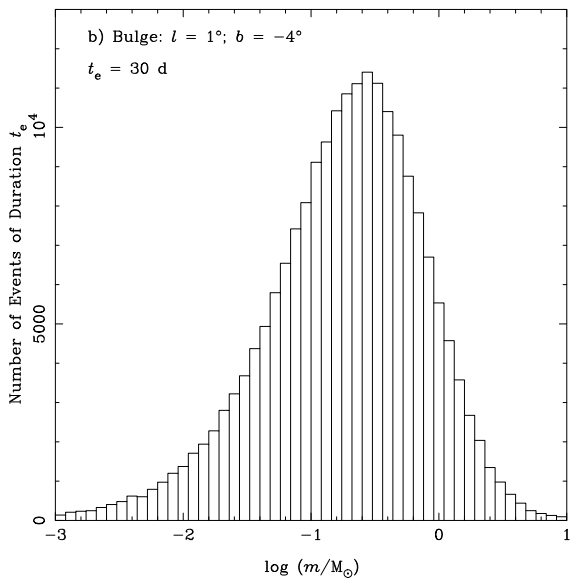

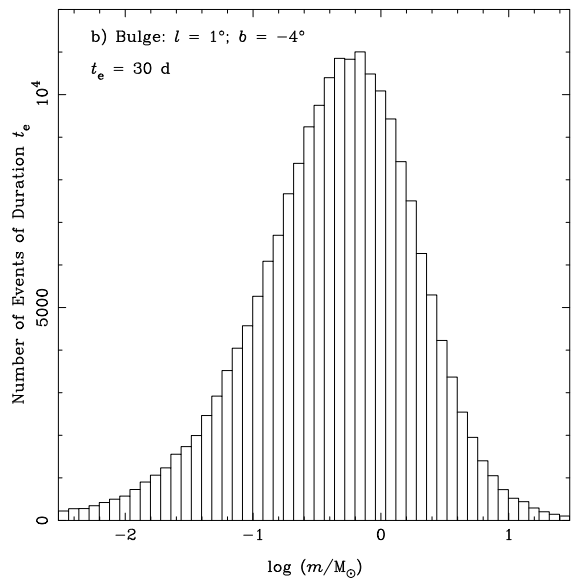

**Figure 3.** The number of simulated thick-disc events of mass $m$ with event duration $t_e$ along the (a) LMC and (b) Galactic bulge (Baade's Window) lines of sight. The most likely lens mass is $0.03\,\mathrm{M}_\odot\,(t_e/15\,\mathrm{d})^2$ for the LMC direction and $0.3\,\mathrm{M}_\odot\,(t_e/30\,\mathrm{d})^2$ for the bulge.

**Figure 4.** The number of simulated bulge events of mass $m$ with event duration $t_e$ along the (a) LMC and (b) Galactic bulge (Baade's Window) lines of sight. The most likely lens mass is $0.03\,\mathrm{M}_\odot\,(t_e/15\,\mathrm{d})^2$ for the LMC direction and $0.5\,\mathrm{M}_\odot\,(t_e/30\,\mathrm{d})^2$ for the bulge.

source distances. The lens mass is given by

$$m = 0.08\ \mathrm{M}_\odot \left(\frac{t_e}{20\ \mathrm{d}}\right)^2 \left(\frac{D}{10\ \mathrm{kpc}}\right)^{-1} \left(\frac{V_t}{220\ \mathrm{km\ s^{-1}}}\right)^2,\tag{3.1}$$

where $D \equiv x(L-x)/L$. The uncertainty is contained in the unknown lens distance $x$ and tangential motion $V_t$ (and also, for observations towards the bulge, the source distance $L$).

Griest (1991) has shown how an estimate of the lens mass can be made from the variation of $d\Gamma/dt_e$ with mass for a given value of $t_e$. De Rújula, Jetzer & Massó (1991) have shown how the moments of the event duration can also be used to extract information on the lens mass function. This method has been used by Jetzer (1994a; 1994b) to estimate the mass of the first LMC and bulge events. In this section, we wish to calculate the lens mass probability distributions for each of the four Galactic components, both towards the LMC and Galactic bulge. In the previous section we used the average observed timescales to indicate likely lens mass ranges and to compare observed and predicted lensing rates. Here, we shall see how well those very approximate mass range estimates correlate with more rigorous statistical determinations. There are numerous theoretical and observational constraints on many of the baryonic dark matter candidates (Carr 1994), and these can be used in conjunction with the lens mass estimates to try to determine the dominant lensing components.

For our lens mass estimates we employ a Monte-Carlo simulation of $2 \times 10^5$ lensing events for each component and each line of sight. In the case of the bulge line of sight a source distance $L$ is selected at random for each event, assuming the distribution of eqn (2.8), with $\beta = 2$ and $\rho_s$ given by eqn (2.7). For the LMC line of sight $L$ is taken to be fixed at 50 kpc. For the lens distance $x$ a random value between 0 and $L$ is selected assuming a probability distribution $P(x) dx = d\tau(x)/\tau(<L) \propto D(x, L)\rho_l(x)$, where $\rho_l$ depends upon the component being considered. A random velocity is then assigned to the lens (and also the source in the case of the bulge line of sight) using eqn (2.5) and the contribution of any systematic motion (i.e. any rotational velocity component of the lens or motion of the line of sight at the lens position) taken into account to give a relative transverse lens velocity $V_t$. For a given timescale the lens mass is then specified by eqn (3.1). For the LMC line of sight we adopt $t_e = 15$ d and for the bulge direction $t_e = 30$ d as being representative of the observed event durations.

The results are shown in Figs. (1–4) for halo, thin disc, thick disc and bulge lenses, respectively. For the LMC direction [Figs. 1(a)–4(a)] the most likely lens mass is 0.05 $M_\odot$, 0.02 $M_\odot$, 0.03 $M_\odot$ and 0.03 $M_\odot$ for the halo, thin disc, thick disc and bulge components, respectively. The implication is that Galactic lenses observed towards the LMC are likely to be brown dwarfs (no matter which component is responsible for the observed events), although the 95% CL mass range is around 2.5 orders of magnitude for the halo and 2 orders of magnitude for the other components, so microlensing observations alone do not rule out M-dwarf stars. However, high-latitude searches for M-dwarfs in the optical and near-infrared (Bahcall, Flynn, Gould & Kirhakos 1994; Hu, Huang, Gilmore & Cowie 1994) have yielded significantly fewer stars than required to explain the observed optical depth. It therefore seems that the LMC microlensing events are due to brown dwarfs.

In the case of the lensing events observed towards Baade's Window [Figs. 1(b)–4(b)], the most likely lens mass is 1.2 $M_\odot$, 0.3 $M_\odot$, 0.3 $M_\odot$ and 0.5 $M_\odot$ for the halo, thin disc, thick disc and bulge components, with a 95% CL range of around 2.5 orders of magnitude in each case. In the case of the halo, a lens mass of 1.2 $M_\odot$ implies either hydrogen-burning stars or stellar remnants. The idea of the dark halo being comprised of ordinary main-sequence stars is clearly at odds with observations, whilst a halo comprised of stellar remnants would over-enrich the interstellar medium with metals, unless the mass function is finely tuned to around 5 $M_\odot$ (Ryu, Olive & Silk 1990). In any case, the disagreement between the halo lens masses as inferred along the LMC and bulge lines of sight implies that the halo is not dominating the rate along at least one of these directions. Timescales of around 10 d towards the bulge are required to provide agreement between the inferred halo lens masses along both lines of sight, and to provide consistency with other observational and theoretical constraints. Since the observed timescales are typically higher than this, the halo cannot be dominating the rate towards the bulge. The observed rate is anyway much higher than can be accommodated by the halo. For the other components M-dwarf or brown dwarf mass scales are preferred. However, optical and near-infrared searches for M-dwarfs imply that they cannot be major contributors to the lensing rate (Bahcall, Flynn, Gould & Kirhakos 1994; Hu, Huang, Gilmore & Cowie 1994), unless the local mass function is unrepresentative of the mass function averaged along the line of sight towards Baade's Window. Similarly, the disc mass function must rise sharply below the hydrogen-burning limit (0.08 $M_\odot$) if brown dwarfs are to provide the required lensing rate in our model (Kroupa, Tout & Gilmore 1993; Kerins & Carr 1994).

## 3.2 Event amplifications

For a peak amplification $A$ the minimum line-of-sight distance of the source star from the lens is smaller than $R_e$ by a factor $u = 2^{1/2}[A(A^2 - 1)^{-1/2} - 1]^{1/2}$ [the inverse of eqn (2.2)]. Hence, one can infer the impact parameter directly from observations.

The probability of a lensing event having an impact parameter in the range $(u, u+du)$ is $P(u) du \propto du$ and so for a large number of events one would expect the mean impact parameter to be $\langle u \rangle = 0.5$ or $A(\langle u \rangle) = 2.18$ for a maximum 'threshold' impact parameter $u_t = 1$. The probability of observing $n$ independent events with impact parameters $u_i$ and mean $\overline{u} = \sum_{i=1}^{n} u_i/n$ relative to $n$ events with a mean $\langle u \rangle = u_t/2$ is

$$P_{\overline{u}}/P_{\langle u \rangle} = \begin{cases} (\overline{u}/\langle u \rangle)^{n-1} & (0 \le \overline{u} \le \langle u \rangle) \\ [(u_t - \overline{u})/\langle u \rangle]^{n-1} & (\langle u \rangle \le \overline{u} \le u_t) \end{cases} . \quad (3.2)$$

Thus, an improbable sample would have $P_{\overline{u}}/P_{\langle u \rangle} \ll 1$. The above likelihood estimator holds true provided there are no observational selection effects which favour some amplifications more than others. Inevitably, however, such selection effects do arise and so eqn (3.2) is as much an estimate of the importance of selection effects as it is a measure of the likelihood of the sample.

The MACHO collaboration adopts a minimum amplification cutoff of $A = 1.5$, implying $u_t = 0.83$ and $\langle u \rangle = 0.42$. The 27 bulge events discussed by Sutherland et al. (1995) have a mean impact parameter $\overline{u} \simeq 0.39$ [from Fig. 4 of Sutherland et al. (1995)]. The relative likelihood of this sample is therefore $P_{\overline{u}}/P_{\langle u \rangle} \simeq 0.3$. This likelihood is extremely sensitive to $\overline{u}$ for a sample as large as 27 events and so the fact that $P_{\overline{u}}/P_{\langle u \rangle}$ is anywhere close to 1 is an indication of a very uniform sample. The MACHO team has detected a further 18 candidates (Bennett et al. 1995) but the amplifications for these await publication. MACHO has also detected 4 LMC events with amplifications 1.52, 1.99, 2.98 and 7.2, giving a mean impact parameter $\overline{u} = 0.47$ and $P_{\overline{u}}/P_{\langle u \rangle} = 0.67$, which again implies a pretty uniform sample. The combined likelihood for the 4 LMC and 27 bulge events is $P_{\overline{u}}/P_{\langle u \rangle} \simeq 0.4$.

The EROS team adopts $u_t = 1.0$, so for its sample $\langle u \rangle = 0.5$. It has so far detected only 2 lensing events from their longer timescale search and these have amplifications of 2.7 and 3.0 giving $\overline{u} = 0.37$ and $P_{\overline{u}}/P_{\langle u \rangle} = 0.75$. This high likelihood, more than anything, reflects the fact that we are dealing with only 2 events. EROS finds no candidate events with $A > 1.2$ from its short timescale CCD observations (Aubourg et al. 1995) although, as discussed in Section 2.3.2, the 5 light curves with $1 < A < 1.2$ could be interpreted as very low mass objects lensing large LMC stars. In this case, the impact parameter distribution would not obey eqn (3.2) (Simmons, Newsam & Willis 1995).

The OGLE team has not adopted an amplification threshold and so very low amplification events are accepted provided their light curves have sufficiently small photometric errors. However, in the absence of selection effects, the impact parameter distribution should still be uniform below some adopted threshold $u_t$. There are now 12 OGLE bulge events, one of which is thought to be a double lens. Of the remaining 11, 3 have $A < 1.34$ and so taking an adopted threshold of $u_t = 1$ leaves 8 events. The mean impact parameter for these 8 is $\overline{u} = 0.35$, implying $P_{\overline{u}}/P_{\langle u \rangle} = 0.08$ from eqn (3.2). This appears to imply that the OGLE sample is quite improbable. However, OGLE has conducted a more sophisticated analysis of its impact parameter distribution (Udalski et al. 1994) which takes account of photometric selection effects. OGLE determines $u_t$ for each event, based on the photometric errors at the time of observation, and takes $u/u_t$, rather than $u$, as an indicator of the uniformity of the sample. $u/u_t$ should be uniformly distributed between 0 and 1 and so $\langle u/u_t \rangle$ should converge to 0.5 for a sufficiently large sample. Eqn (3.2) can then be used to test the distribution of $u/u_t$ rather than $u$. For the 9 single-lens events reported by Udalski et al. (1994), $\overline{u/u_t} = 0.53$ and so $P_{\overline{u/u_t}}/P_{\langle u/u_t \rangle} = 0.58$, which indicates that the sample is indeed consistent with the microlensing hypothesis.

# 4   CONCLUSIONS

We have investigated the expected lensing properties of a 'standard Galaxy' model with spherically-symmetric halo and bulge/spheroid components. The simplifying assumption of spherical-symmetry is naive and, in the case of the bulge, is not favoured by direct observations, but we are here interested in the extent to which microlensing rules out such a model.

Using the observed event durations towards the LMC we estimate the total rate due to dark matter to be around 16 events per year for MACHO observations, most of which comes from the dark halo. MACHO detected only 3 events in its first year and so, if one takes the contribution of the disc and bulge components to be fixed, this limits the halo fraction in compact objects to $f_h < 0.4$ at the 95% CL. This limit is $f_h < 0.34$ if the bulge is dominated by dark matter. However, a 'no halo compact matter' hypothesis is ruled out at the 80% CL or greater unless one assumes that the LMC itself has a substantial halo comprised of such objects (De Rújula, Giudice, Mollerach & Roulet 1994). A scenario in which the LMC halo is comprised of compact objects, but our own Galactic halo is not, seems rather contrived.

We have also computed lens-mass probability distributions for each component, based on the observed event durations, and have found that, for the LMC direction, the lens is likely to be a brown dwarf, irrespective of which component the lens actually resides. Whilst M-dwarfs are not ruled out by microlensing observations, their contribution is strongly constrained by near-infrared searches (Bahcall, Flynn, Gould & Kirhakos 1994; Hu, Huang, Gilmore & Cowie 1994).

For the bulge line of sight we find an expected lensing rate of around $2-5$ events per year for OGLE observations (again using the observed timescales as a guide and counting as events only those detections for which $u < 1$), which is comparable to what they are observing. The predicted optical depth is a factor 2 or 3 less than the current observational determinations but these estimates are based on only 10 events (or 13 in the case of MACHO) and so this discrepancy is probably not too serious. More detections are required to conclusively show up any discrepancy between spherically-symmetric and asymmetric bulge models.

Lensing mass determinations for the bulge line of sight point to M-dwarf or brown dwarf mass scales in the case of the disc or bulge. To provide consistency with near-infrared limits one needs to assume that the lenses are brown dwarfs rather than M-dwarfs and, even then, one requires a sharp turn-up in the slope of the mass function below the hydrogen-burning limit to provide the required rate, unless the local mass function is unrepresentative of the mass function averaged along the bulge line of sight. For lenses in the halo component a mass scale around $1~M_\odot$ is preferred on the basis of the observed event durations. Such objects could not be hydrogen-burning and so must necessarily be stellar remnants. However, the possible contribution of stellar remnants is already strongly constrained by ISM metallicity observations and so the implication is that the halo cannot be dominating the rate towards the bulge. This is consistent with a halo only partially comprised of compact matter, as has already been inferred from LMC microlensing observations.

A naive analysis of the distribution of impact parameters for the EROS, MACHO and OGLE candidates, which ignores possible photometric bias, shows that the EROS and MACHO candidates are quite consistent with the microlensing hypothesis. The OGLE sample is also consistent with microlensing provided careful account is taken of photometric effects. This is primarily because the OGLE collaboration has a different selection procedure from MACHO and EROS in that it does not adopt an amplification cutoff.

The EROS team has recently placed strong limits on the density contribution of very low mass ($10^{-8}~M_\odot < m < 10^{-2}~M_\odot$) objects on the basis that its short timescale CCD search has uncovered no lensing events with $A > 1.2$. However, EROS does have 5 low-amplification light curves which might be best explained as the microlensing of large source stars by very low mass lenses. In such a case the point-source amplification breaks down and so one would expect only low-amplification events to occur. On the basis of recent work on finite-size source lensing by Simmons, Newsam & Willis (1995) we estimate that such low-amplification events are extremely common for lens masses approaching $10^{-7}~M_\odot$. Follow-up observations of the source stars themselves should help to discriminate between this suggestion and the possibility of intrinsic variability.

## ACKNOWLEDGMENTS

I wish to thank Bernard Carr and the anonymous referee for helpful comments and advice, and the UK Particle Physics and Astronomy Research Council for receipt of a research studentship.